\documentclass[prb,floatfix,twocolumn,showpacs,amsmath,amssymb]{revtex4}

\usepackage{graphicx}% Include figure files
\usepackage{dcolumn}% Align table columns on decimal point
\usepackage{bm}% bold math

\begin{document}
\title{
Coherence Peak in the Spin Susceptibility 
from Nesting in Spin-Triplet Superconductors;
A Probe for  Line Nodes in Sr$_2$RuO$_4$
%Dynamical susceptibility of the spin-triplet superconductivity in
%Sr$_2$RuO$_4$ 
}

\author{Mayumi Yakiyama and Yasumasa Hasegawa}
%\email{hasegawa@sci.himeji-tech.ac.jp}
\affiliation{
Department of Material Science,
Graduate School of Science,\\
Himeji Institute of Technology\\
 Ako, Hyogo 678-1297, Japan}
%\homepage{http://}

\date{\today}

%\draft

\begin{abstract}
We study 
the dynamical spin susceptibility $\chi ({\bf q},\omega)$ for
spin-triplet superconductivity.
We show that a large peak at $\omega = 2 \Delta$ 
appears in $\textrm{Im} 
\chi_{zz}({\mathbf{Q}}, \omega)$, 
where $z$ is the direction of the $\mathbf{d}$ vector for triplet pairing,
if Fermi surface has a nested part 
with the nesting vector $\mathbf{Q}$ 
and the order parameter is $+\Delta$ and $-\Delta$ 
in this part of the Fermi surface.
If there are line nodes in the nested part of the Fermi surface,
a peak appears in either   $\textrm{Im} \chi_{zz}(\mathbf{Q}, \omega)$ or 
 $\textrm{Im} \chi_{+-}(\mathbf{Q}, \omega)$, or both, 
 depending on the perpendicular 
component of the nesting vector.
The comparison with inelastic neutron scattering experiments  can determine
the position of the line nodes in triplet superconductor 
Sr$_2$RuO$_4$.

\end{abstract}
\pacs{
%PACS numbers: 
74.25 Ha, 74.70 Pq, 61.12 Bt
}

\maketitle
%\hspace{0.5cm}

%\narrowtext
\section{Introduction}

The superconductivity in
Sr$_2$RuO$_4$\cite{Maeno94} has been revealed to be unconventional
by many experiments;
it is spin-triplet\cite{Ishida98,Duffy2000},
it breaks time-reversal symmetry\cite{Luke98},
and  its energy gap has 
line nodes\cite{Ishida2000,Nishizaki2000,Lupien2001,Suzuki2002}. 

%$\mathbf{d}(\mathbf{k})=\Delta (\sin a k_x/2 \cos a k_y/2
%+ i \cos ak_x/2 \sin ak_y/2) \cos c k_z/2$,
%where $a$ and $c$ are the lattice constants, 
The triplet superconductivity with
 horizontal line nodes of the energy gap %on the Fermi surface
has been proposed\cite{Hasegawa2000} to explain these
experiments.
The absence of the angle dependence of the 
thermal conductivity
 within the $a$-$b$ plane \cite{Tanatar2001,Izawa2001}
 shows that the line nodes run horizontally on the
Fermi surface.
The angle resolved ultrasound attenuation experiment\cite{Lupien2001} 
is compatible
with the horizontal line nodes.

The Fermi surface of  Sr$_2$RuO$_4$  consists of three cylindrical
surfaces named $\alpha$, $\beta$ and
$\gamma$\cite{Mackenzie96,Damascelli2000},
as predicted by 
 the band calculation\cite{Oguchi95,Singh95}. 
The hybridization of  the $d_{xz}$
and $d_{yz}$ orbits of ruthenium 
with the $p_{\pi}$ orbit of oxygen makes  two one-dimensional bands, 
if the mixing of these bands is neglected.
If  small mixing is taken into account, we get 
hole-like $\alpha$ and electron-like $\beta$, but
the  nesting  of the  Fermi surfaces survives
as predicted\cite{Mazin99} and
confirmed by the inelastic neutron scattering experiment\cite{Sidis99}.
 The $\gamma$ surface is
constructed by the $d_{xy}$ orbit of Ruthenium and is two-dimensional. 

Nomura and Yamada\cite{Nomura2002} 
have studied the two-dimensional three-band Hubbard model in the 
third-order perturbation theory
and obtained that triplet superconductivity is stabilized 
mainly in the two-dimensional band. They also obtained 
 the line-node-like power-law behavior in the temperature-dependence of
the specific heat due to the vertical node-like energy gap
 in the $\alpha$ and $\beta$ bands. 
%%__added_2002/9/26__%%%%
In their treatment the momentum dependence of the gap is determined at 
the transition temperature $T_c$.
The spin-triplet state such as
\begin{equation}
{\bf d}({\bf k})=\hat{\bf z} 
  \Delta \sin  k_x \sin  k_y (\sin  k_x + i \sin  k_y),
\end{equation}
or
\begin{equation}
{\bf d}({\bf k})=\hat{\bf z} 
  \Delta (\cos^2  k_x - \cos^2  k_y) (\sin  k_x + i \sin  k_y),
\end{equation}
where ${\bf d}({\bf k})$ is the ${\bf d}$ vector, 
should be mixed with 
\begin{equation}
\hat{\bf z} \Delta' (\sin  k_y + i \sin  k_x),
\end{equation}
or
\begin{equation}
\hat{\bf z} \Delta' (\sin  k_x - i \sin  k_y),
\end{equation}
at $T<T_c$
and the vertical line nodes will disappear\cite{Hasegawa2000,Zhitomirsky2001}.

 Zhitomirsky and Rice\cite{Zhitomirsky2001} 
proposed the mechanism for the horizontal line nodes that
while the active band has a full energy gap,
 the interband
proximity effect makes the horizontal line nodes in the passive band.
They assumed that the
two-dimensional band is  active and that the line nodes are
in one-dimensional bands. 

On the other hand,
It has been shown that spin-triplet superconductivity is
induced in the quasi-one dimensional bands by the
antiferromagnetic spin fluctuation if the antiferromagnetic spin
fluctuation is anisotropic in the spin space\cite{Sato2000,Kuwabara2000}.
 If this is the case the
one-dimensional bands are expected to have a full energy gap
and the two-dimensional band has line nodes.

Therefore, the determination of the gap structure 
is important to understand the mechanism of unconventional superconductivity 
in Sr$_2$RuO$_4$.
Bulk measurements such as specific heat, NMR and thermal
conductivity cannot distinguish which 
part of the Fermi surface has  the line nodes,
$\alpha$, $\beta$ or $\gamma$. 

In the inelastic neutron scattering experiments, 
imaginary part of the dynamical spin susceptibility, Im $\chi({\bf q}, \omega)$,
is observed, from which we can get the information of the superconducting
order parameter.
The so-called 41 meV peak in the spin-singlet $d$-wave superconducting state of
YBa$_2$Cu$_3$O$_7$\cite{RossatMignod91,Mook93,Fong95,Bourges96,Dai2001}
has been observed. Many theoretical 
studies\cite{Lu92,Tanamoto93,Tanamoto94,Maki94,Bulut93,Marsiglio93,Lavagna 94,Demler95,Liu95,Mazin95,Brinckmann99}
have been done to explain the
peak structure in inelastic neutron scattering.

Dynamical susceptibility for the spin-triplet superconductivity
has been studied theoretically\cite{Brinkman74,Joynt88}.
Recently, the resonance peak in 
$\textrm{Im} \chi (\mathbf{q},\omega)$ is shown to be a sign 
of  the triplet superconductivity\cite{Morr2001,Kee2000,Fay2000}.
The order parameter assumed in these papers, however, 
does not seem  to be consistent with experiments in Sr$_2$RuO$_4$.

In this paper we show the general form of
  the dynamical spin  susceptibility
$\chi_{ij}(\mathbf{q},\omega)$ in the unitary states of 
the triplet superconductivity
and calculate it in the system which has the nested Fermi surface
with and without line nodes in the nested part of the Fermi surface.
 We show that
the position of the line nodes can be determined
by the inelastic neutron scattering experiment, which observes the imaginary
part of the dynamical susceptibility\cite{Sidis99,Servant2002,Braden2002}.

%%%%%%%%%%%%%%%%%%%%%%%%%%%%%%%%%%%%%%%%%%%%%%%%%%%%%%%%%%%%%%%%%%%%%%%%%%%
\section{dynamical spin susceptibility of spin-triplet superconductivity}
The dynamical spin susceptibility is given 
by\cite{Brinkman74,Joynt88,Morr2001,Kee2000}
\begin{eqnarray}
 & &\chi_{ij}^0(\mathbf{q},i\omega_m) 
 = - \frac{1}{4} 
\nonumber \\
 & & \times T \sum_{n,\mathbf{k}}
  {\rm Tr} \left( \hat{\alpha}_i \hat{G} (\mathbf{k}, i\epsilon_n) 
  \hat{\alpha}_j \hat{G} (\mathbf{k} + \mathbf{q}, i \epsilon_n+i\omega _m)
  \right),
\label{eqchi}
\end{eqnarray}
where $\omega_m=2 m \pi T$  and $\epsilon_n=(2 n +1) \pi T$ are  
Matsubara frequencies 
($m$ and $n$ are integers), and
 $\hat{\alpha}$ and $\hat{G}(\mathbf{k}, i\epsilon_n)$ are the $4
\times 4$ Nambu representation of the spin and Green function,
respectively, i.e., 
\begin{eqnarray}
 \hat{\alpha}_i &=& \left(
 \begin{array}{cc}
    \sigma^i & 0 \\
    0        & \sigma^y \sigma^i \sigma^y
 \end{array}
 \right) 
 \nonumber \\ 
 &=& \frac{1+\rho_z}{2} \sigma^i 
    + \frac{1-\rho_z}{2}\sigma^y \sigma^i\sigma^y
\end{eqnarray} 
where $\sigma^i$ ($i=x$, $y$, or $z$) is a Pauli matrix, and
\begin{equation}
 \hat{G}(\mathbf{k}, i \epsilon_n)=
  \left(
    \begin{array}{ll}
   G(\mathbf{k},i\epsilon_n)           & F(\mathbf{k},i\epsilon_n) \\
   F^{\dagger}(\mathbf{k},i\epsilon_n) & -G(-\mathbf{k},-i\epsilon_n)
    \end{array} \right).   
\end{equation}
The $2 \times 2$ matrix
 Green function  $G(\mathbf{k},i\epsilon_n)$  and the anomalous Green
function $F(\mathbf{k},i\epsilon_n)$ are 
 given as the Fourier coefficients for 
\begin{equation}
  G_{\alpha, \beta}(\mathbf{k}, \tau) = 
  -\langle {\rm T}_\tau a_{\mathbf{k} \alpha}(\tau) 
 a^\dagger_{\mathbf{k}
     \beta}(0) \rangle 
\end{equation}
and
\begin{equation}
  F_{\alpha, \beta}(\mathbf{k}, \tau) = 
  \langle {\rm T}_\tau a_{\mathbf{k} \alpha}(\tau) a_{-\mathbf{k}
     \beta}(0) \rangle\\,
\end{equation}
respectively.

In this paper we consider the weak coupling theory for the spin-triplet
superconductivity. 
We  take  account of  the interaction $U$ by
the random phase approximation (RPA), 
\begin{equation}
 \chi_{ij}({\bf q},\omega)=
 \frac{\chi_{ij}^{0}({\bf q}, \omega)}{1-U \chi_{ij}^{0}({\bf q}, \omega)} ,
\end{equation}
but the essential 
properties such as a peak in $\textrm{Im} \chi_{ii}({\bf q},\omega)$
is already seen in the absence of the interaction effects.

The order parameter is given by the $\mathbf{d}$ vector
as
%\begin{equation}
$\Delta_{\alpha \beta}(\mathbf{k})=
 i ( (\mathbf{d}(\mathbf{k}) \cdot {\bm \sigma}) \sigma^y
    )_{\alpha \beta}$.
%\end{equation}
We study  the unitary states, $\mathbf{d}^*(\mathbf{k}) 
\times \mathbf{d}(\mathbf{k}) =0$,
  in this paper, 
since experimental results can be explained by the unitary states.
%we do not have to consider much complicated non-unitary states. 
%Then the Green
%function and the anomalous Green function are given in the simple
%forms;
%\begin{equation}
%   G_{\alpha, \beta}(\mathbf{k}, i \epsilon_n) =
%\frac{-\delta_{\alpha, \beta}(i\epsilon_n + \xi_\mathbf{k})}
%     {\epsilon_n^2 + E_\mathbf{k}^2},    
%\end{equation} 
%and
%\begin{equation}
%   F_{\alpha, \beta}(\mathbf{k}, i \epsilon_n) =
%\frac{\Delta_{\alpha, \beta}}
%     {\epsilon_n^2 + E_\mathbf{k}^2},   
%\end{equation}
%where 
%\begin{equation}
% E_\mathbf{k} =\sqrt{\xi_\mathbf{k}^2 + |\mathbf{d}(\mathbf{k})|^2}.
%\end{equation} 
Then we perform the summation over $n$ 
in Eq.(\ref{eqchi}), and get
\begin{eqnarray}
& &  \chi_{ij}^{0}(\mathbf{q},\omega) =
\frac{1}{4} \sum_{\alpha \alpha' \beta \beta'}
 \sigma^i_{\alpha \alpha'} \sigma^j_{\beta \beta'}
 \nonumber \\
&\times& \sum_\mathbf{k} \bigg\{
  C^{(+)}_{\alpha \alpha' \beta \beta'}
 (\mathbf{k},\mathbf{q}) D^{(-)}(\mathbf{k},\mathbf{q},\omega)
  \big( f(E_{\mathbf{k}'}) -  f(E_\mathbf{k}) \big)  
\nonumber \\
 &+&
  C^{(-)}_{\alpha \alpha' \beta \beta'}
 (\mathbf{k}, \mathbf{q}) D^{(+)}(\mathbf{k},\mathbf{q},\omega)
  \big( 1- f(E_{\mathbf{k}'}) -  f(E_{\mathbf{k}}) \big) \bigg\},
\nonumber \\
\label{eqchiq2}
\end{eqnarray}
where
\begin{eqnarray}
& &
  C^{(\pm)}_{\alpha \alpha' \beta \beta'}(\mathbf{k},\mathbf{q})
  =
    \frac{\delta_{\alpha \beta'} \delta_{\alpha' \beta}}{2}
 \nonumber \\
   &\pm& 
    \frac{\delta_{\alpha \beta'} \delta_{\alpha' \beta}\xi_\mathbf{k}
    \xi_{\mathbf{k}'}
   - \textrm{Re} ( \Delta_{\alpha \beta}^* (\mathbf{k})
    \Delta_{\alpha' \beta'}(\mathbf{k}') ) }
    {2 E_{\mathbf{k}}E_{\mathbf{k}'}},
\end{eqnarray}
\begin{eqnarray}
  D^{(\pm)}(\mathbf{k},\mathbf{q},\omega)
 &=&
  \frac{1}{E_\mathbf{k} \pm E_{\mathbf{k}'}+ \omega + i \Gamma} 
 \nonumber \\
 &+& 
      \frac{1}{E_{\mathbf{k}} \pm E_{\mathbf{k}'} - \omega - i\Gamma} ,
 \end{eqnarray}
 $\mathbf{k}' = \mathbf{k} + \mathbf{q}$,
and the analytic continuation $i \omega_m \rightarrow \omega + i \Gamma$
with $\Gamma \rightarrow +0$ has been done.

%For the spin-triplet superconductivity we get
%\begin{eqnarray}
%& &\sum_{\alpha \alpha' \beta \beta'} \sigma^i_{\alpha \alpha'}
% \sigma^j_{\beta \beta'}
%\Delta^*_{\alpha \beta}(\mathbf{k})
% \Delta_{\alpha' \beta'}(\mathbf{k}')=
%2 (\delta_{i j} \mathbf{d}^*(\mathbf{k})\cdot \mathbf{d}(\mathbf{k}')
% \nonumber \\
% &-& d_i^*(\mathbf{k}) d_j (\mathbf{k}') 
% - d_j^*(\mathbf{k}) d_i (\mathbf{k}'))
%\end{eqnarray} 
%\begin{eqnarray}
% & &\sum_{\alpha \alpha' \beta \beta'} \sigma^i_{\alpha \alpha'}
% \sigma^j_{\beta \beta'}
%\Delta^*_{\alpha \beta}(\mathbf{k})
% \Delta_{\alpha' \beta'}(\mathbf{k}')
%\nonumber \\
%&=&
%\left\{ \begin{array}{ll}
%   2 d_z^*(\mathbf{k}) d_z (\mathbf{k}') & \mbox{ if $i=j=x$ or $i=j=y$} \\
% - 2 d_z^*(\mathbf{k}) d_z (\mathbf{k}') & \mbox{ if $i=j=z$ }\\
%   0                               & \mbox{ otherwise.}
%  \end{array} \right.
%\end{eqnarray}
%Then we get 
We take $\mathbf d$ vector parallel to the $z$ axis as indicated by 
 experiments\cite{Ishida98,Duffy2000}, to obtain
$\chi_{ij}^{0}=0$ for $i \not= j$ and
\begin{eqnarray}
& &\chi_{ii}^{0}(\mathbf{q},\omega)
\nonumber \\
&=& \frac{1}{2} \sum_\mathbf{k}\bigg\{
\tilde{C}_{ii}^{(+)}
 (\mathbf{k},\mathbf{q})D^{(-)}(\mathbf{k},\mathbf{q})\left(
f(E_{\mathbf{k}'}) - f(E_\mathbf{k})\right)
\nonumber \\
&+& \tilde{C}_{ii}^{(-)}
 (\mathbf{k},\mathbf{q})D^{(+)}(\mathbf{k},\mathbf{q})\left(
1-f(E_{\mathbf{k}'}) - f(E_\mathbf{k})\right)\bigg\},
\end{eqnarray}
where
\begin{eqnarray}
\tilde{C}_{xx}^{(\pm)}(\mathbf{k},\mathbf{q})
 &=&
\tilde{C}_{yy}^{(\pm)}(\mathbf{k},\mathbf{q})
=\tilde{C}_{+-}^{(\pm)}(\mathbf{k},\mathbf{q})
\nonumber\\
&=& \frac{1}{2} \pm 
\frac{\xi_\mathbf{k}\xi_{\mathbf{k}'}
 - \textrm{Re}(d_z^*(\mathbf{k})d_z(\mathbf{k}'))}
{2 E_\mathbf{k} E_{\mathbf{k}'}},
\label{eqcxx} 
\end{eqnarray}
and
\begin{eqnarray}
\tilde{C}_{zz}^{(\pm)}(\mathbf{k},\mathbf{q})&=&
 \frac{1}{2} \pm 
\frac{\xi_\mathbf{k}\xi_{\mathbf{k}'}
 + \textrm{Re} (d_z^*(\mathbf{k})d_z(\mathbf{k}'))}
{2 E_\mathbf{k} E_{\mathbf{k}'}}.
\label{eqczz}
\end{eqnarray}

Depending on the sign of $\textrm{Re} (d^*_z(\mathbf{k}) d_z(\mathbf{k}'))$ in 
the coherence factor, $\tilde{C}_{ii}^{(-)}$ in 
Eqs. (\ref{eqcxx}) and (\ref{eqczz}),   a peak will appear in
either $\textrm{Im} \chi_{+-}(\mathbf{q},\omega)$ or 
$\textrm{Im} \chi_{zz}(\mathbf{q},\omega)$, i.e.
if   $\textrm{Re} (d^*_z(\mathbf{k}) d_z(\mathbf{k}')) >0$ ($<0$), 
a coherence peak  appears in $\chi_{+-}(\mathbf{q}, \omega)$
 ($\chi_{zz}(\mathbf{q}, \omega)$), as we will show in
the next section.

\section{coherence peak from the nested Fermi surface}
\begin{figure}[tb]
\includegraphics[width=0.45\textwidth]{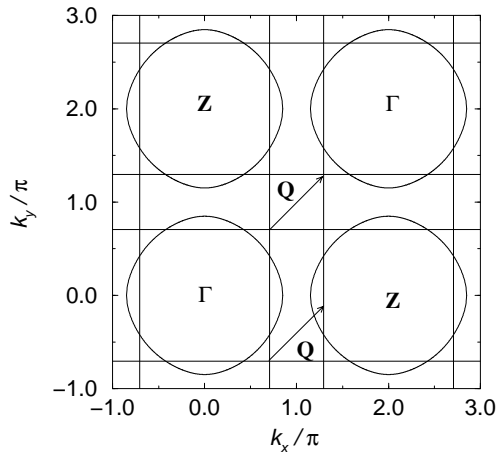}
%  \begin{center}
  %% \mbox{\psfig{figure=fig1.eps,width=7cm}}
  %% \epsfile{file=fig1.eps,width=7cm}
%  \leavevmode \epsfxsize=8.5cm  \epsfbox{fermis.eps}
%  \end{center}
 \caption{Fermi surface at $k_z=0$
and the nesting vector $\mathbf{Q}$.
}
\label{figfermi}
\end{figure}

In order to study the coherence
peak in Sr$_2$RuO$_4$, we take the simple 
three-band model, where  there are two  
one-dimensional bands 
and a two-dimensional band, i.e., 
\begin{eqnarray}
 \epsilon_\mathbf{k}^{(1)}&=&-2 t_{1} \cos( k_x) + \epsilon_{01} 
   \label{1dband}\\
 \epsilon_\mathbf{k}^{(2)}&=&-2 t_{1} \cos( k_y) + \epsilon_{01}\\
 \epsilon_\mathbf{k}^{(3)}&=&-2 t_{3} (\cos(k_x) +\cos(k_y))\nonumber\\
  & &-4 t_{3}' \cos(k_x) \cos(k_y) + \epsilon_{03} ,
\end{eqnarray}
where the lattice constant is taken to be $1$.
We set parameters as
$t_{1}=0.31$ eV, $\epsilon_{01}=-0.24$ eV, $t_{3}=0.44$ eV, $t_{3}'=0.14$ eV
and $\epsilon_{03}=-0.14$ eV. 
%In the RPA we take $U=0.175$ meV.
The Fermi surface is  
 shown in Fig. \ref{figfermi}.
%
%%%%%%%%%%%%%%%%%%%%%%%%%%%%%%%%%%%%%%%%%%%%%%%%%%%%%%%%%%%%%%%%%%%%%%%
%  
%%%%%%
%Due to the perfect nesting of the one-dimensional 
%Fermi surface ($\xi_\mathbf{k}=-\xi_{\mathbf{k}'}$),
%$Re \chi(\mathbf{Q}, 0)$ for the normal state diverges logarithmically
% as $T$ goes to zero,
%where $\mathbf{Q} \approx (0.7 \pi, 0.7 \pi, Q_z)$ is the nesting vector
%as shown in Fig. \ref{figfermi}.
%Even if the nesting is imperfect and the mixing 
%of the two one-dimensional bands are 
%taken into account,
%this quasi-one dimensional property is not changed as 
%predicted\cite{Mazin99} and observed in the inelastic neutron scattering 
%in the normal state\cite{Sidis99}. 
We set the wave vector $\bf q$ to be   the nesting vector $\mathbf Q$. 
The $z$ component of $\mathbf Q$ is arbitrary if $\mathbf{d}(\mathbf{k})$
does not depend on $k_z$. Since
 $\textrm{Im} \chi_{ii}^{0}(\mathbf{Q},\omega)$ 
is  dominated by the contribution from the one 
dimensional bands,
we consider the dynamical spin-susceptibility only 
for the one-dimensional bands in this paper.

For the normal state with perfectly nested Fermi 
surface ($\xi_{\mathbf{k}'}=-\xi_\mathbf{k}$)
the imaginary part of the dynamical susceptibility is
\begin{equation}
\textrm{Im} \chi_{\textrm{normal}}^{0}
(\mathbf{Q},\omega)=\frac{\pi}{2} N(0) \left(
1-2 f\left( \frac{\omega}{2} \right) \right),
\end{equation}
where $f(\omega/2)$ is the Fermi distribution function and 
the constant density of states 
$N(0)$ is assumed.

We study  three possible cases for the triplet superconductivity. 
In the first case (case A) 
the constant energy gap 
opens in the one-dimensional bands, while
the line nodes are in the two-dimensional band.
In the second case (case B) 
we assume that the order parameter
in the one-dimensional bands depends only on $k_z$  
as $\cos k_z$.
%and
%the full gap on the two-dimensional Fermi surface.
In the last case (case C) the order parameter
in the one-dimensional bands  depends both on $k_y$ and $k_z$
%on the one-dimensional Fermi surface
and it is zero in the horizontal line nodes.

%In the RPA we take $\Delta_1=0.001$, $T=0.0001$,  $\Gamma=0.0001$
%and $U=0.175$. 

%%%%%%%%%%%%%%%%%%%%%%%%%%%%%%%%%%%%%%%%%%%%%%%%%%%%%%%%%%%%%%%%
\subsection{case A}
First we study the case A, where we take the order parameters as
\begin{eqnarray}
 d_{1z}(\mathbf{k})&=&   \Delta_1 \sin k_x 
\label{eqA1} \\
 d_{2z}(\mathbf{k})&=& i \Delta_1 \sin k_y ,
\label{eqA2}
\end{eqnarray}
and
\begin{eqnarray}
d_{3z}(\mathbf{k})&=&\Delta_3 \big(\sin\frac{k_x}{2} \cos\frac{k_x}{2}
 +i\cos\frac{k_x}{2}\sin\frac{k_y}{2}\big) 
  \nonumber \\
 & & \times
\cos\frac{k_z}{2} .
\end{eqnarray}
In the case A, Cooper pairs are formed between 
the electrons on the nearest sites 
in the conducting plane ($\mathbf{r}_i$ and 
$\mathbf{r}_i + (\pm a,0,0)$, $\mathbf{r}_i + (0,\pm a,0)$)
 for the one-dimensional electrons, and Cooper pairs are formed  
between the electrons on $\mathbf{r}_i$ and $\mathbf{r}_i 
+ (\pm a/2, \pm b/2, \pm c/2)$  for the two-dimensional electrons.
%The size of the Cooper pairs or the coherence length can be larger 
%than the lattice size
%as is the case in the usual spin-singlet $s$ wave superconductivity, 
%which is caused by the 
%effective on-site attraction in the tight-binding electrons.
These order parameters can be realized if 
the pairing Hamiltonian is taken as\cite{Zhitomirsky2001}
\begin{eqnarray}
 {\cal{H}}'&=& \sum_{\mathbf{k},\mathbf{k}',\sigma,\sigma'} \bigg( g_{11}  \big\{
\sin k_x \sin k_x'
a^{\dagger}_{\mathbf{k}\sigma}  a^{\dagger}_{-\mathbf{k} -\sigma}
a_{-\mathbf{k}'-\sigma'}  a_{\mathbf{k} \sigma'} \nonumber \\
&+&
 \sin k_y \sin k_y'
b^{\dagger}_{\mathbf{k}\sigma}  b^{\dagger}_{-\mathbf{k} -\sigma}
b_{-\mathbf{k}'-\sigma'}  b_{\mathbf{k} \sigma'}  \big\} 
\nonumber \\
&+& g_{13}
\big\{ 
 \sin k_x 
 \sin \frac{k_x'}{2}  \cos \frac{k_y'}{2} \cos \frac{k_z'}{2}  
\nonumber \\
&\times&
(a^{\dagger}_{\mathbf{k}\sigma}a^{\dagger}_{-\mathbf{k}-\sigma}
c_{-\mathbf{k}'-\sigma'}c_{\mathbf{k}'\sigma'} + h.c.)
\nonumber \\ 
&+& 
 \sin k_y
  \cos \frac{k_x'}{2} \sin \frac{k_y'}{2} \cos \frac{k_z'}{2} 
\nonumber \\
&\times&
(b^{\dagger}_{\mathbf{k}\sigma}b^{\dagger}_{-\mathbf{k}-\sigma}
c_{-\mathbf{k}'-\sigma'}c_{\mathbf{k}'\sigma'} + h.c.) \big\} \bigg),
\end{eqnarray} 
where $a^{\dagger}_{\mathbf{k} \sigma}$ 
and $b^{\dagger}_{\mathbf{k} \sigma}$,
 are creation operators for the
electron in one-dimensional bands 
and  $c^{\dagger}_{\mathbf{k} \sigma}$ is the 
creation operator for the two-dimensional band.
In the above  only terms relevant to the spin-triplet superconductivity with
$\mathbf{d}(\mathbf{k}) \parallel {\hat{\mathbf{z}}}$ are included. 
The terms proportional to $g_{11}$ describe
 the attractive interaction between electrons with up and down spins
 in the one-dimensional bands which
makes the triplet superconductivity with $\mathbf d$ 
vector parallel to the $z$ axis.
The $g_{13}$ terms represent the pair hoppings between 
one-dimensional and two-dimensional Fermi surface.
%
%This type of superconductivity is possible when the one-dimensional
%bands are active and the two-dimensional band is passive.

\begin{figure}[tb]
\includegraphics[width=0.45\textwidth]{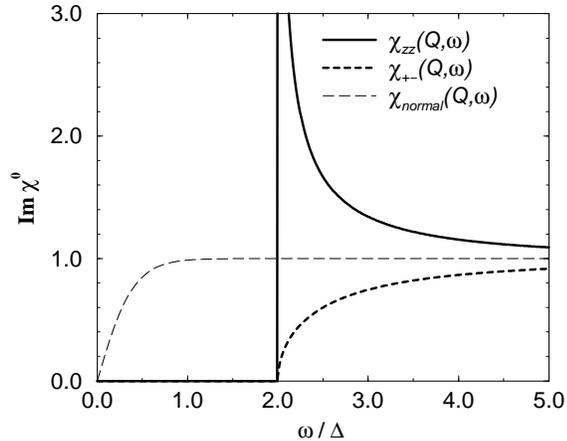}
 \caption{Imaginary part of the dynamical susceptibility
normalized by $(\pi/2) N(0)$
in case A (full gap on one-dimensional Fermi surface, eqs. (\ref{eqA1}) 
and (\ref{eqA2})).
We take $T=0$ for the superconducting state
and we take finite $T$ for the normal state. 
(Since $\omega$ and $T$ is scaled by $\Delta$, we take  $T=0.1 \Delta$
for the normal state.)}
\label{figcaseA0}
\end{figure}

In this case $\textrm{Im} \chi_{ii}^{0}(\mathbf{Q},\omega)$ 
from the one-dimensional band
%at $T=0$ 
is obtained   as
\begin{eqnarray}
\textrm{Im} \chi_{zz}^{0}(\mathbf{Q}, \omega) &=&
\left\{
\begin{array}{ll}
 0 & \mbox{if ${\omega} < 2 \Delta$} \\
 \frac{\pi}{2} N(0) \frac{\omega}{\sqrt{
{\omega}^2-(2 \Delta)^2}} \tanh \frac{\omega}{4 T}
 & \mbox{if ${\omega} \geq 2 \Delta$}
\end{array} \right. \nonumber \\ \\
\textrm{Im} \chi_{+-}^{0}(\mathbf{Q}, \omega) &=&
\left\{
\begin{array}{ll}
 0 & \mbox{if ${\omega} < 2 \Delta$}  \\
 \frac{\pi}{2} N(0) \frac{\sqrt{\omega^2-(2\Delta)^2}}
{\omega}  \tanh \frac{\omega}{4 T}
 & \mbox{if $\omega \geq 2 \Delta$}
\end{array} \right.  ,
\nonumber  \\ 
\end{eqnarray} 
where $\Delta=\Delta_1 \sin k_F$ is the energy gap 
on the Fermi surface
in the one-dimensional bands.
%We calculate the $\omega$ dependence of
%${\textrm Im} \chi_{ii}(\mathbf{Q}, \omega)$.
%As seen in Fig. \ref{figcaseA}, resonance peak 
%at $\omega \approx 2 \Delta_1$
%appears in ${\textrm Im} \chi_{zz}$ while no resonance peak appears 
%in ${\textrm Im} \chi_{+-}$.
%%%%%%%%%%%%%%%%%%%%%%%%%%%%%%%%%%%%%%%%%%%%%%%%%%%%%%%%%%%%%%%%%%%%%%%
%\begin{figure}[tbh]
%\includegraphics[width=0.45\textwidth]{caseA.eps}
% \caption{Imaginary part of the dynamical susceptibility
%in case A (full gap in one-dimensional Fermi surface)
%for $q$ being the nesting vector $\mathbf{q}_{\parallel}=\mathbf{Q}$.
%}
%\label{figcaseA}
%\end{figure}
%%%%%%%%%%%%%%%%%%%%%%%%%%%%%%%%%%%%%%%%%%%%%%%%%%%%%
In Fig. \ref{figcaseA0}
we plot imaginary part of the dynamical susceptibility
normalized by $(\pi/2)N(0)$ 
 as a function of $\omega$
at $T=0$
for the superconducting state and at a
 finite temperature for the normal state ($T=0.1\Delta$). 
Since the coherence factor ${\tilde C}^{(-)}_{zz}({\bf k},{\bf Q})$ is $1$,
$\textrm{Im} \chi_{zz}^{0}({\bf Q},\omega)$ diverges at $\omega=2\Delta$
as the density of states in the $s$-wave superconductivity.

\begin{figure}[tb]
\includegraphics[width=0.45\textwidth]{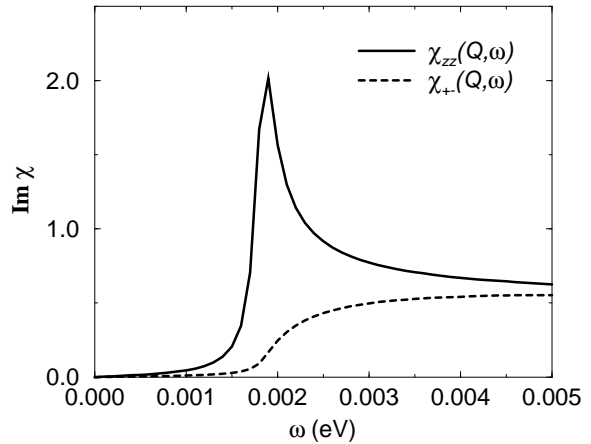}
 \caption{Imaginary part of the dynamical susceptibility
in case A (full gap on one-dimensional Fermi surface, eqs. (\ref{eqA1}) 
and (\ref{eqA2})) calculated in RPA.
We take $\Delta_1=0.001$ eV, $T=0.0001$ eV,  $\Gamma=0.0001$ eV
and $U=0.175$ eV.}
\label{figcaseARPA}
\end{figure}

In Fig.~\ref{figcaseARPA} we plot the dynamical susceptibility obtained in RPA
with parameters, $\Delta_1=0.001$ eV, $T=0.0001$ eV,  $\Gamma=0.0001$ eV
and $U=0.175$ eV for the one-dimensional band (Eq.~(\ref{1dband})
and $\sin k_F \approx 0.922$).  
A peak appears at $\omega = 2 \Delta_1 \sin k_F \approx 0.00184$ eV 
only in $\textrm{Im} \chi_{zz}({\bf Q}, \omega)$.

%
%For the normal state we take $T=0.1\Delta$.
%
%%%%%%%%%%%%%%%%%%%%%%%%%%%%%%%%%%%%%%%%%%%%%%%%%%%%%%%%%%%%%%%%%%%%%%%
%%%%%%%%%%%%%%%%%%%%%%%%%%%%%%%%%%%%%%%%%%%%%%%%%%%%%
%%%%%%%%%%%%%%%%%%%%%%%%%%%%%%%%%%%%%%%%%%%%%%%%%%%%%%%%%%%%%%%%%%%%%%%
%%%%%%%%%%%%%%%%%%%%%%%%%%%%%%%%%%%%%%%%%%%%%%%%%%%%%

%%%%%%%%%%%%%%%%%%%%%%%%%%%%%%%%%%%%%%%%%%%%%%%
\subsection{case B}
\begin{figure}[tb]
\includegraphics[width=0.45\textwidth]{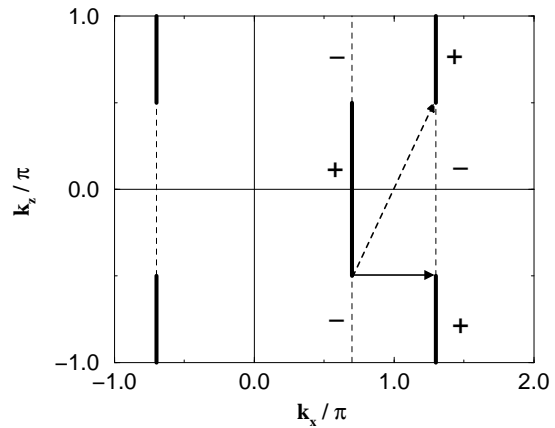}
\caption{
Nesting vector $\mathbf Q$ in the case B  
(horizontal line nodes on one-dimensional 
Fermi surface,
eqs. (\ref{eqB1}) and (\ref{eqB2}) ). 
Thick solid and thin dashed lines show the side view of the 
Fermi surface on which order parameter is  
positive and negative, respectively.
Solid arrows ($Q_z=0$) and dashed arrows ($Q_z=\pi$)
are the nesting vectors. 
}
\label{figqz}
\end{figure}

%%%%%%%%%%%%%%%%%%%%%%%%%%%%%%%%%%
Next we study the case B. 
We assume horizontal line nodes on the one-dimensional Fermi surface as
\begin{eqnarray}
d_{1z}(\mathbf{k})&=&   \Delta_1 \sin k_x \cos k_z 
\label{eqB1} 
\\
d_{2z}(\mathbf{k})&=& i \Delta_1 \sin k_y \cos k_z
\label{eqB2}
\end{eqnarray}
In this case the Cooper pairs in the one-dimensional bands are formed
between electrons on ${\bf r}$ and ${\bf r} + (0,0,\pm c)$.
Although this order parameter is not likely to be realized in Sr$_2$RuO$_4$,
we study this case to show the mechanism of resonance peak.
In this case if we take $Q_z=0$ or $\pi$, line nodes are connected by 
$\mathbf Q$ as shown in 
Fig. \ref{figqz}.
%%%%%%%%%%%%%%%%%%%%%%%%%%%%%%%%%%%%%%%%%%%%%%%%%%%%%%%%%%%%%%%%%%%%%%%
%%%%%%%%%%%%%%%%%%%%%%%%%%%%%%%%%%%%%%%%%%%%%%%%%%%%%
%
%%%%%%%%%%%%%%%%%%%%%%%%%%%%%%%%%%%%%%%%%%%%%%%%%%%%%%%%%%%%%%%%%%%%%%%
\begin{figure}[tb]
\includegraphics[width=0.45\textwidth]{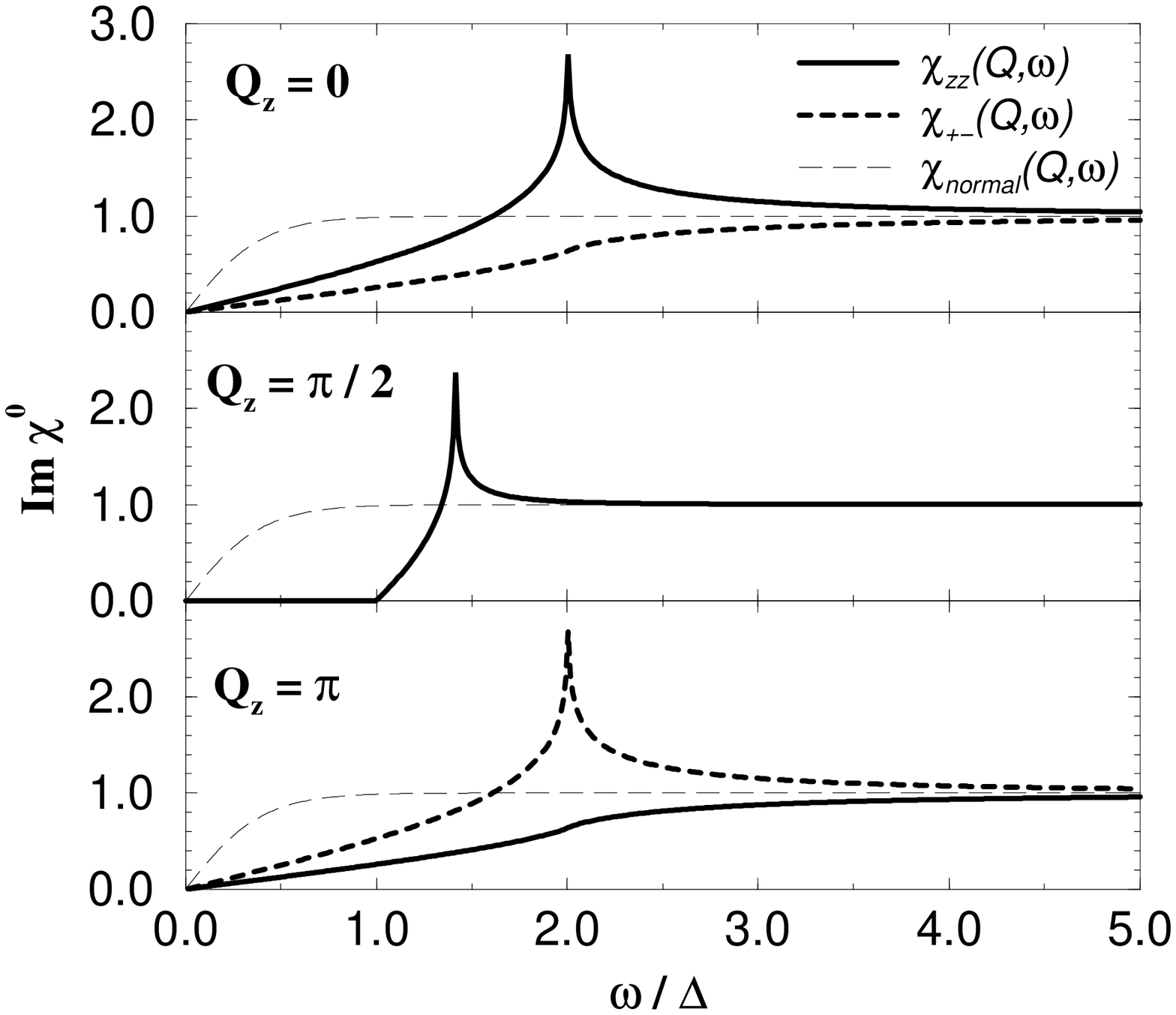}
 \caption{Imaginary part of the dynamical susceptibility
in the case B (line nodes in one-dimensional Fermi surface
eqs. (\ref{eqB1}) and (\ref{eqB2}) ).
We take the same parameters as in Fig.~\ref{figcaseA0}.
}
\label{figcaseB}
\end{figure}
%%%%%%%%%%%%%%%%%%%%%%%%%%%%%%%%%%%%%%%%%%%%%%%%%%%%%
%
For $Q_z=0$ we obtain  that 
\begin{eqnarray}
& &\textrm{Im} \chi_{zz}^{0}(\mathbf{Q}, \omega) \nonumber \\
&=&
\left\{
\begin{array}{ll}
 N(0) F\left(\arcsin\left( \frac{\omega}{2\Delta}\right) |
 \left(\frac{2\Delta}{\omega} \right)^2 \right) \tanh \frac{\omega}{4T} & \\
 &  \hspace{-3cm} 
  \mbox{if } {\omega} < 2 \Delta  \\
  N(0) K \left( \left( \frac{2\Delta}{\omega}\right)^2\right)
    \tanh \frac{\omega}{4T} & \\
 &  \hspace{-3cm} 
  \mbox{if } {\omega} \geq 2 \Delta
\end{array} \right.  
\nonumber \\ \\
& &\textrm{Im} \chi_{+-}^{0}(\mathbf{Q}, \omega) \nonumber \\
&=&
\left\{
\begin{array}{ll}
 N(0) E\left( \arcsin \left(\frac{\omega}{2\Delta}\right) |
\left(\frac{2\Delta}{\omega}\right)^2 \right) \tanh \frac{\omega}{4T} & \\
 &  \hspace{-3cm} 
  \mbox{if } {\omega} < 2 \Delta  \\
 N(0) E\left(\left(\frac{2\Delta}{\omega}\right)^2\right) \tanh \frac{\omega}{4T}
  & \\
  & \hspace{-3cm} 
   \mbox{if } \omega \geq 2 \Delta
\end{array} \right.  ,
\nonumber \\
\end{eqnarray}
where $F(\phi|m)$  and $E(\phi|m)$ are incomplete elliptic integrals of 
the first and the second kinds, and $K(m)$ and $E(m)$ are 
complete elliptic integrals of 
the first and second kinds, respectively.
As shown in the top panel of Fig. \ref{figcaseB},
$\textrm{Im} \chi_{zz}^{0}({\bf Q}, \omega)$
 diverges at $\omega=2\Delta$
as the density of states for superconductivity with line nodes.
For $Q_z=\pi$, $\textrm{Im} \chi_{zz}^{0}$ 
and $\textrm{Im} \chi_{+-}^{0}$ are exchanged.

For $Q_z=\pi/2$ we get at $T=0$
\begin{eqnarray}
& &
\textrm{Im} \chi_{zz}^{0}(\mathbf{Q}, \omega) = 
\textrm{Im} \chi_{+-}^{0}(\mathbf{Q},\omega) \nonumber \\
&=&\left\{
  \begin{array}{ll}
 0 & \hspace{-3cm} \mbox{if $\omega < \Delta$} \\
 N(0) F \left(
 \arcsin ((\frac{\omega}{\Delta})^2-1) | \frac{1}{((\omega/\Delta)^2-1)^2}
 \right)
 & \\
  & \hspace{-3cm}
 \mbox{if $\Delta \leq \omega < \sqrt{2}\Delta$} \\
 K \left( \frac{1}{((\omega/\Delta)^2-1)^2}\right)  & \\
 & \hspace{-3cm}
 \mbox{if $\omega \geq \sqrt{2} \Delta$}.
 \end{array} \right.
\end{eqnarray}

\begin{figure}[tb]
\includegraphics[width=0.45\textwidth]{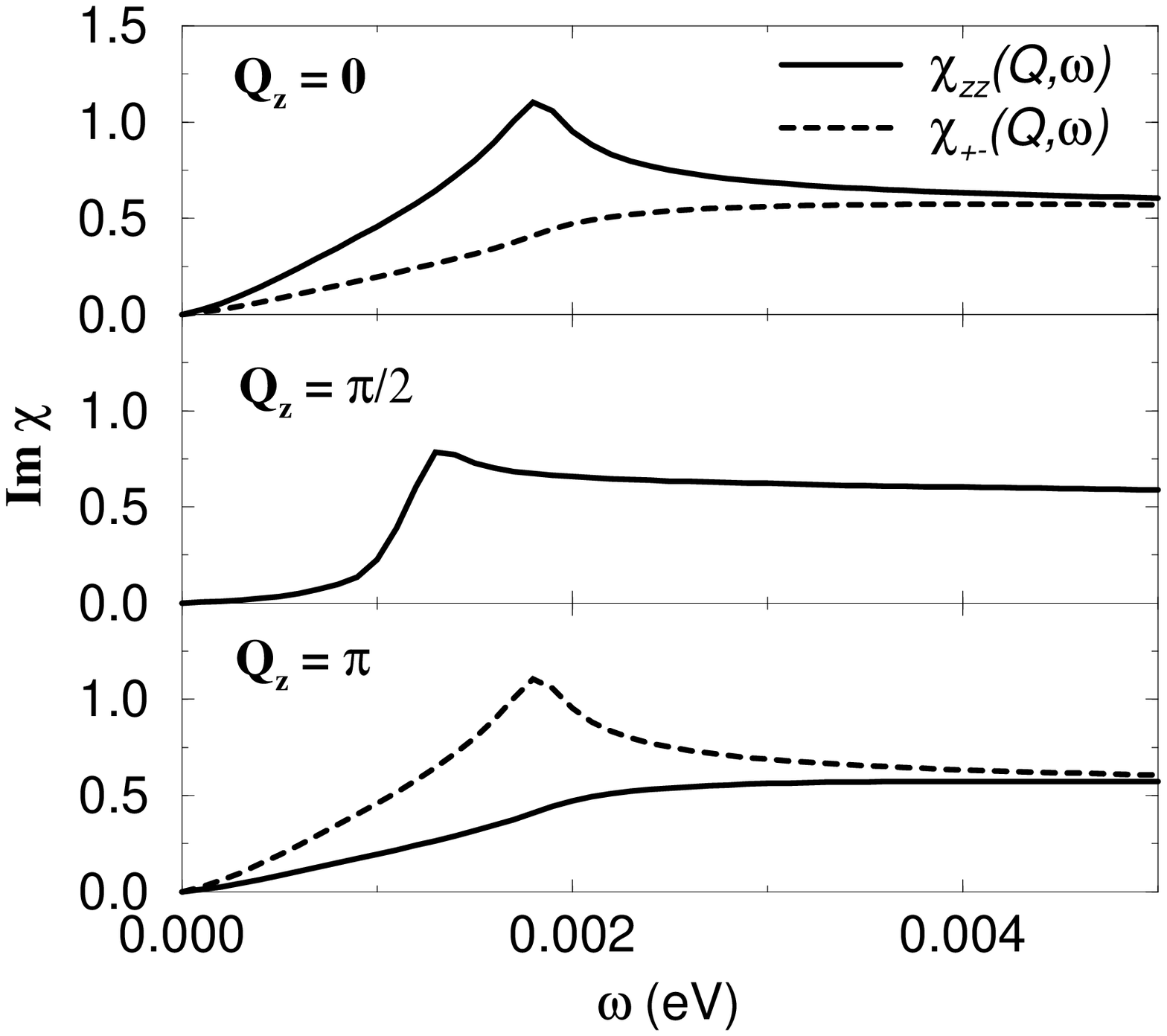}
 \caption{Imaginary part of the dynamical susceptibility
in the case B (line nodes in one-dimensional Fermi surface
eqs. (\ref{eqB1}) and (\ref{eqB2}) ) calculated in RPA.
We take 
the same parameters as in Fig.~\ref{figcaseARPA}.}
\label{figcaseBRPA}
\end{figure}

In the RPA we get essentially the same 
result as shown in Fig.~\ref{figcaseBRPA}.

\begin{figure}[tb]
\includegraphics[width=0.45\textwidth]{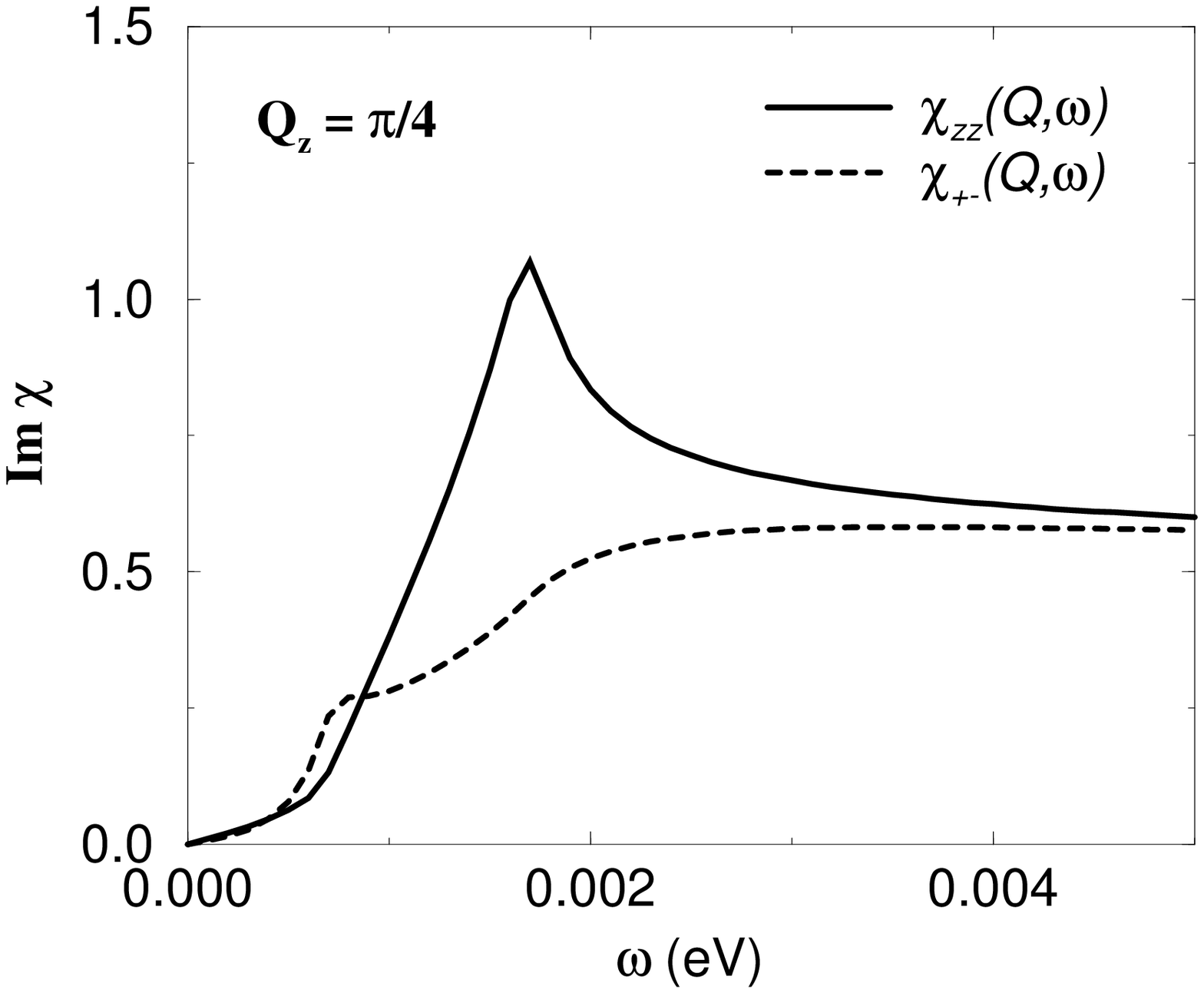}
 \caption{Imaginary part of the dynamical susceptibility
in the case B (line nodes in one-dimensional Fermi surface
eqs. (\ref{eqB1}) and (\ref{eqB2}) ) calculated in RPA.
We take the same parameters as in Fig.~\ref{figcaseARPA}.
}
\label{figcaseBRPA2}
\end{figure}

For $Q_z \neq 0$, $\pi/2$, and $\pi$, $|d_{1z}({\bf k})| + |d_{1z}({\bf k}')|$
at $k_z=k_F$
has two local maximum of $2 \Delta_1 \sin k_F |\cos (Q_z/2)|$ and
$2 \Delta_1 \sin k_F |\sin (Q_z/2)|$. 
If
$\textrm{Re} (d_{1z}^{*}({\bf k}) d_{1z}({\bf k}'))>0$  at $k_z$, where 
$|d_{1z}({\bf k})| + |d_{1z}({\bf k}')|$ becomes local maximum,
then a peak appears  in $\textrm{Im}\chi_{+-}({\bf Q},\omega)$, else
 a peak appears  in $\textrm{Im}\chi_{zz}({\bf Q},\omega)$.
In Fig.~\ref{figcaseBRPA2} we plot the RPA result for $Q=\pi/4$.
A peak is seen at $\omega=2 \Delta_1 \cos \pi/8
\approx 0.0017$ in $\textrm{Im}\chi_{zz}({\bf Q},\omega)$ and
a small peak  is seen at  $\omega=2 \Delta_1 \cos \pi/8
\approx 0.0007$ in  $\textrm{Im}\chi_{+-}({\bf Q},\omega)$.

%%%%%%%%%%%%%%%%%%%%%%%%%%%%%%%%%%%%%%%%%%%%%%%%%%%%%%%%%%%%%%%%%%%%%%%
%%%%%%%%%%%%%%%%%%%%%%%%%%%%%%%%%%%%%%%%%%%%%%%%%%%%%
%
%%%%%%%%%%%%%%%%%%%%%%%%%%%%%%%%%%%%%%%%%%%%%%%%%%%%%%%%%%%%%%%%%%%%%%%
%%%%%%%%%%%%%%%%%%%%%%%%%%%%%%%%%%%%%%%%%%%%%%%%%%%%%
%

%%%%%%%%%%%%%%%%%%%%%%%%%%%%%%%%%%%%%%%
\subsection{case C}
\begin{figure}[tb]
\includegraphics[width=0.45\textwidth]{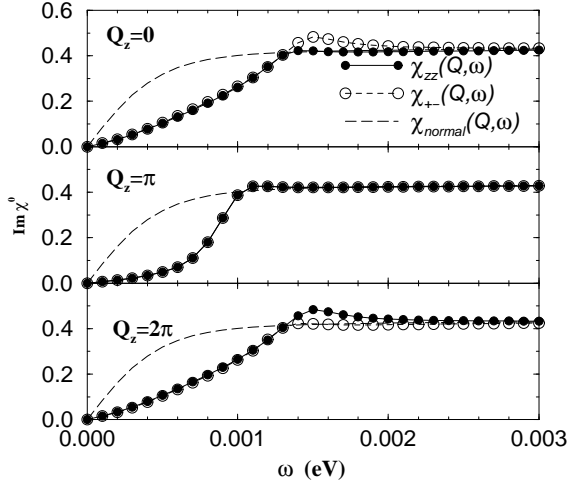}
 \caption{Imaginary part of the dynamical susceptibility
in the case C (line nodes on one-dimensional Fermi surface,
eq.~(\ref{eqcaseC})).
We take $\Delta_1=0.001$ eV, $T=0.0001$ eV,  $\Gamma=0.0001$ eV.
}
\label{figcaseC}
\end{figure}

%%%%%%%%%%%%%%%%%%%%%%%%%%%%%
Finally we study the case C.
Here we assume the order parameter 
in the one-dimensional bands 
to be
\begin{eqnarray}
& & d_{1z}(\mathbf{k})=d_{2z}(\mathbf{k}) \nonumber \\
 &=&
 \Delta_1 \left( \sin \frac{k_x}{2} \cos \frac{k_y}{2}
 + i \cos \frac{k_x}{2} \sin \frac{k_y}{2}
 \right)
 \nonumber \\  &\times&
\cos \frac{k_z}{2}
\label{eqcaseC}
\end{eqnarray} 
This type of superconductivity is realized if the 
two-dimensional band is active and the one-dimensional bands are
passive\cite{Zhitomirsky2001}. 
In this case the $\Gamma$ and $Z$ points in Fig.\ref{figfermi}
are not equivalent and
the order parameter is zero at $k_z=\pm \pi$, $\pm 3 \pi$.
We calculate $\textrm{Im} \chi_{ii}(\mathbf{Q},\omega)$ numerically 
%by taking 
%the one dimensional band as
%$\epsilon_\mathbf{k}^{(1)}=-2 t_{1} \cos(a k_x) + \epsilon_{01}$
%with 
%$t_{1}=0.31$ eV, $\epsilon_{01}=-0.24$ eV,
%$\Delta_1=1$ meV, $T=0.1$ meV, and $\Gamma=0.1$ meV. 
 and plot $\textrm{Im} 
\chi_{ii}(\mathbf{Q}, \omega)$ as a function of  $\omega$ 
in Fig. \ref{figcaseC}.
%
%%%%%%%%%%%%%%%%%%%%%%%%%%%%%%%%%%%%%%%%%%%%%%%%%%%%%%%%%%%%%%%%%%%%%%%%
%%%%%%%%%%%%%%%%%%%%%%%%%%%%%%%%%%%%%%%%%%%%%%%%%%%%%%%%%%%%%%%%%%%%%%%%
\begin{figure}[tb]
\includegraphics[width=0.45\textwidth]{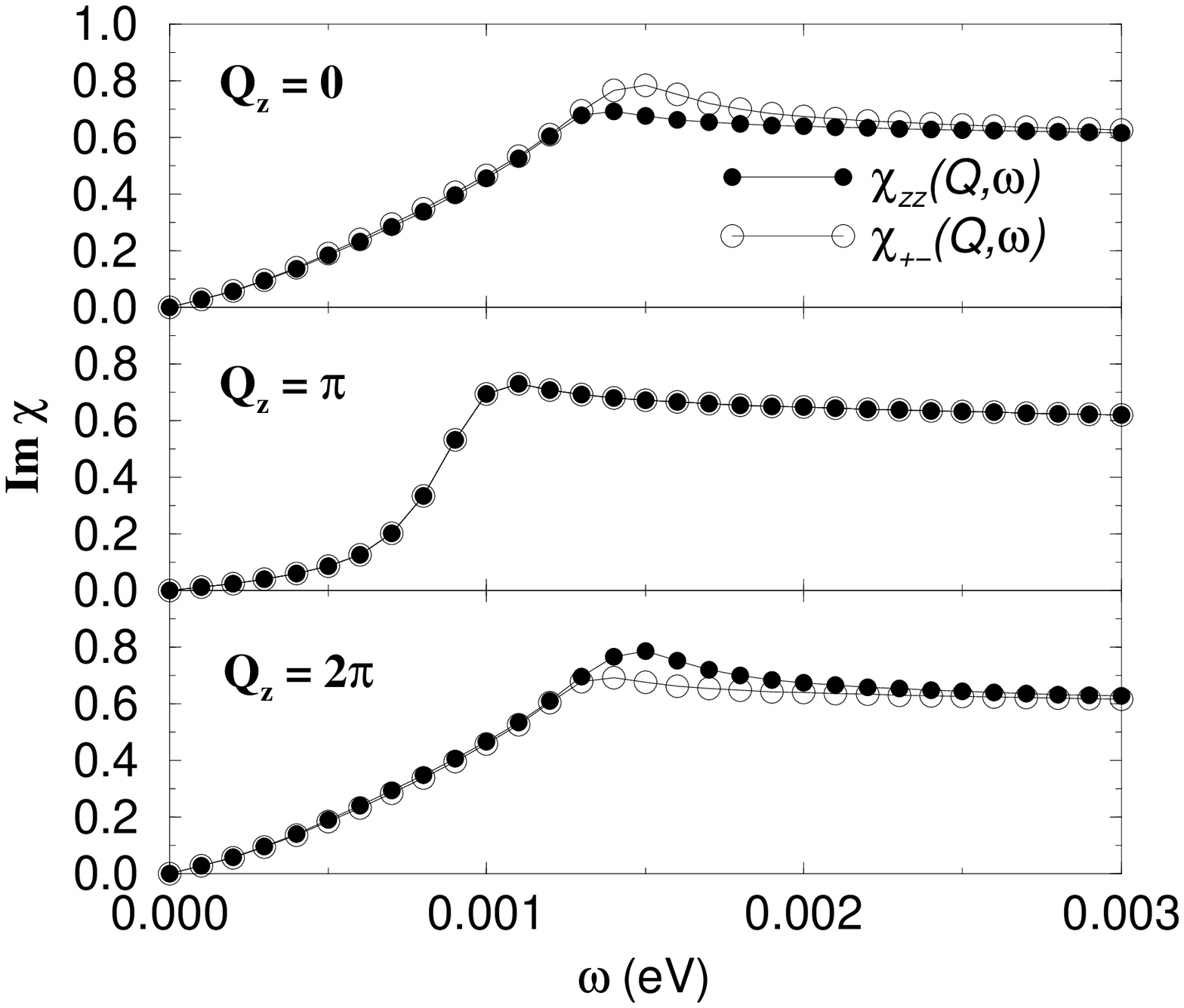}
 \caption{Imaginary part of the dynamical susceptibility
in the case C (line nodes on one-dimensional Fermi surface,
eq.~(\ref{eqcaseC}))
calculated in RPA.
We take 
the same parameters as in Fig.~\ref{figcaseARPA}.
}
\label{figcaseCRPA}
\end{figure}
%
%%%%%%%%%%%%%%%%%%%%%%%%%%%%%%%%%%%%%%%%%%%%%%%%%%%%%
%%%%%%%%%%%%%%%%%%%%%%%%%%%%%%%%%%%%%%%%%%%%%%%%%%%%% 
%For $Q_z=0$ and $2 \pi$, 
%a linear dependence on $\omega$
%is seen in  $\omega  \Delta_1$, since particle-hole 
%excitations around the nodes 
%are possible as in the case B. For $q_z=\pi$, on the other hand,
%the gap structure is seen 
%in the $\omega$ dependence in  ${\textrm Im} \chi_{ii}$. 
%%%%%%%%%%%%%%%%%%%%%%%%%%%%%%%%%%%%%%%%%%%%%%%%%%%%%%%%%%%%%%%%%%%%%%%
%\begin{figure}[tbh]
%\includegraphics[width=0.45\textwidth]{figqz0pi.eps}
% \caption{Imaginary part of the dynamical susceptibility
%in case B (line nodes in one-dimensional Fermi surface).
%}
%\label{figqz0}
%\end{figure}
%%%%%%%%%%%%%%%%%%%%%%%%%%%%%%%%%%%%%%%%%%%%%%%%%%%%%
The  small coherence peak exists at $\omega \approx 1.5  \Delta_1$.
In this case 
$\textrm{Re} (d^*_z(\mathbf{k}) d_z(\mathbf{k}+\mathbf{q}))$ 
changes sign for $\mathbf k$ 
on the Fermi surface.  
The small coherence peak in $\chi_{+-}(\mathbf{Q}, \omega)$ for $Q_z=0$ 
and that in $\chi_{zz}(\mathbf{Q}, \omega)$ for $Q_z=2 \pi$ can be understood 
as follows. 
%In the one-dimensional band 
%\begin{equation}
% k_F = \cos^{-1} \frac{\epsilon_{01}}{2 t_1}. 
%\end{equation}
At $k_x = k_F \approx 0.63 \pi$ and ${\bf Q} = (2 \pi-2k_F, 2\pi-2k_F, Q_z)$,
the maximum value of
\begin{eqnarray}
& & |d_{1z}({\bf k})| +|d_{1z}({\bf k}')| \nonumber \\
&=&
\frac{\Delta_1}{2} 
 \left( |\cos \frac{k_z}{2}| \sqrt{1-\cos k_F \cos k_y} \right.\nonumber \\
& &
 + \left. |\cos \frac{ k_z+Q_z}{2}| \sqrt{1-\cos k_F \cos ( k_y-2 k_F) }
 \right),
\label{deltakQ}
\end{eqnarray}
is approximately  $1.52 \Delta_1$ 
at $k_y \approx 1.62 \pi$ and $k_z=0$, if $Q_z = 0$ or $2\pi$. 
Since $\textrm{Re} (d_{1z}^*({\bf k}) d_{1z}({\bf k}')) > 0$ ($<0$) 
at  $k_y \approx 1.62 \pi$ and $k_z=0$
for $Q_z=0$ ($Q_z=2\pi$), $\textrm{Im} \chi_{+- (zz)}({\bf Q},\omega)$
has a peak at $\omega \approx 1.52 \Delta_1$.

%%%%%%%%%%%%%%%%%%%%%%%%%%%%%%%%%%%%%%%%%%%%%%%%%%%%%%%%%%%%%%%%%%
\section{comparison with spin-singlet superconductivity}
%%%%%%%%%%%%%%%%%%%%%%%%%%%%%%%%%%%%%%%%%%%%%%%%%%%%%%%%%%%%%%%%%%
Eq. (\ref{eqchiq2}) can be also applied  for the spin-singlet
 case, when the order parameter is written as $\Delta_{\alpha
 \beta}(\mathbf{k})=i \sigma^y_{\alpha \beta} \Delta(\mathbf{k})$.
As expected, we get the isotropic 
 coherence factor
%$\sum_{\alpha \alpha' \beta \beta'} \sigma^i_{\alpha \alpha'}
% \sigma^j_{\beta \beta'}$ 
%$\Delta^*_{\alpha \beta}(\mathbf{k})
% \Delta_{\alpha' \beta'}(\mathbf{k}')
%= -2 \delta_{i,j}$ $\Delta^*(\mathbf{k})\Delta(\mathbf{k}')$. 
\begin{eqnarray}
 & &
\sum_{\alpha \alpha' \beta \beta'} \sigma^i_{\alpha \alpha'}
\sigma^j_{\beta \beta'}
C^{(\pm)}_{\alpha \alpha' \beta \beta'}
\nonumber \\
&=&\delta_{ij}
\left( 1 \pm \frac{\xi_\mathbf{k} \xi_{\mathbf{k}'} + 
  \textrm{Re} ( \Delta^*(\mathbf{k}) \Delta (\mathbf{k}'))}
  {E_\mathbf{k}E_{\mathbf{k}'}}
\right) .
\end{eqnarray}
A peak appears in 
$\textrm{Im} \chi_{ii}(\mathbf{q}, \omega)$,
only if $\textrm{Re} (\Delta^*(\mathbf{k}) \Delta (\mathbf{k}'))<0$,
which is the case 
for the $d$-wave paring with ${\bf q} \approx (\pi, \pi, q_z)$
in high $T_c$ cuprates\cite{Lu92,Lavagna94,Tanamoto94,Maki94,Fong95,Mazin95}.
Note that the dynamical spin-susceptibility
$\chi_{+-}^{0}({\bf q}, \omega)$ (eq.~(\ref{eqcxx})) 
for the spin-triplet superconductivity is similar to 
the dynamical \textit{charge}-susceptibility 
in the spin-singlet superconductivity\cite{Marsiglio93a}.

%%%%%%%%%%%%%%%%%%%%%%%%%%%%%%%%%%%%%%%%%%%%%%%%%%%%%%%%%%%%%%%%%%%%%
\section{conclusion}
We study the dynamical spin-susceptibility for the spin-triplet superconductivity.
The resonance peak appears either in $\textrm{Im} \chi_{zz}({\bf q},\omega)$
or $\textrm{Im} \chi_{+-}({\bf q},\omega)$ 
from the nesting part of the Fermi surface.
We have shown that the existence of the line nodes on 
the quasi-one dimensional Fermi surface drastically changes 
the dynamical susceptibility,
which can be observed by inelastic neutron scattering experiments.

Coherence peak appears in 
$\textrm{Im} \chi_{+-}(\mathbf{Q}, \omega)$ if 
$\textrm{Re}(d^*_z(\mathbf{k}) d_z(\mathbf{k}+\mathbf{Q}))>0$ or
 in $\textrm{Im} \chi_{zz}(\mathbf{Q}, \omega)$ if 
$\textrm{Re}(d^*_z(\mathbf{k}) d_z(\mathbf{k}+\mathbf{Q}))<0$.
When $\textrm{Re}(d^*_z(\mathbf{k}) d_z(\mathbf{k}+\mathbf{Q}))$ changes
sign in the nested part of the Fermi surface,
a coherence peak can  appear
in both  $\textrm{Im} \chi_{zz}(\mathbf{Q}, \omega)$ and
 $\textrm{Im} \chi_{+-}(\mathbf{Q}, \omega)$,
but the divergence becomes weaker and easily smeared out. 
The position of the line nodes on the Fermi surface, if 
line nodes exist in the nested part of the
Fermi surface, will be observed by scanning $Q_z$ in the inelastic 
neutron scattering.  

Recently, the resonance peak was searched in Sr$_2$RuO$_4$ by 
inelastic neutron scattering\cite{Servant2002,Braden2002},
but no changes have been observed below $T_c$ yet.
We take the amplitude of 
the order parameter in the one-dimensional band 
as $\Delta_1=0.001$ eV in the RPA calculation. 
If the one-dimensional band is passive, the value may be smaller
and the observation will be difficult.
The gap structure, however, 
will be observed by experiments
with better resolutions using better single crystals.

Recently,
Mukuda et al\cite{Mukuda2002} reported a different dependence of the relaxation in
Ruthenium and Oxide nuclear quadrupole resonance. 
This difference may be explained by the 
difference of $\chi_{+-}({\bf q}, \omega)$ and $\chi_{zz}({\bf q}, \omega)$
in the spin triplet superconductivity.

%\vspace{-0.2cm}
%


\begin{thebibliography}{99}
%\vspace{-0.2cm}

\bibitem{Maeno94}
Y. Maeno et al.
Y. Maeno, H. Hashimoto,K. Yoshida, S. Nishizaki, T. Fujita,J.G. Bednorz, and
F. Lichitenberg,
Nature (London) \textbf{372}, 532 (1994).

\bibitem{Ishida98}
 K. Ishida, H. Mukuda, Y. Kitaoka, K. Asayama, Z. Q. Mao, Y. Mori and Y. Maeno,
 Nature (London) \textbf{396}, 658 (1998).
%
\bibitem{Duffy2000}
J. A. Duffy, S. M. Hayden, Y. Maeno, Z. Mao, J. Kulda, and G. J. McIntyre,
 Phys. Rev. Lett. \textbf{85}, 5412, (2000).
%
\bibitem{Luke98}
G. M. Luke, Y. Fudamoto, K. M. Kojima, M. I. Larkin, J. Merrin, B. Nachumi, 
Y. J. Uemura, Y. Maeno, Z. Q. Mao, Y. Mori, H. Nakamura, and M. Sigrist,
 Nature (London) \textbf{394}, 558 (1998).

\bibitem{Ishida2000}
K. Ishida, H. Mukuda, Y. Kitaoka, Z. Q. Mao, Y. Mori, and Y. Maeno
Phys. Rev. Lett. \textbf{84}, 5387 (2000).

\bibitem{Nishizaki2000}
S. NishiZaki et al., 
S. NishiZaki, Y. Maeno and Z. Q. Mao,
J. Phys. Soc. Jpn. \textbf{69}, 572 (2000).


\bibitem{Lupien2001}
C. Lupien, W. A. MacFarlane, Cyril Proust, and Louis Taillefer
Phys. Rev. Lett. \textbf{86}, 5986 (2001).


\bibitem{Suzuki2002}
M. Suzuki, M. A. Tanatar, N. Kikugawa, Z. Q. Mao, Y. Maeno, and T. Ishiguro,
Phys. Rev. Lett. \textbf{88}, 227004 (2002).


\bibitem{Hasegawa2000}
Y. Hasegawa, K. Machida, and M. Ozaki, J. Phys. Soc. Jpn. \textbf{69},
336 (2000).

\bibitem{Izawa2001}
K. Izawa, H. Takahashi, H. Yamaguchi, Y. Matsuda, M. Suzuki, T. Sasaki,
 T. Fukase, Y. Yoshida, R. Settai, and Y. Onuki,
 Phys. Rev. Lett. \textbf{86}, 2653 (2001).

\bibitem{Tanatar2001}
M. A. Tanatar, M. Suzuki, S. Nagai, Z. Q. Mao, Y. Maeno, and T. Ishiguro,
Phys. Rev. Lett. \textbf{86}, 2649 (2001).


\bibitem{Mackenzie96}
A. P. Mackenzie, S. R. Julian, A. J. Diver, G. J. McMullan, M. P. Ray,
 G. G. Lonzarich, Y. Maeno, S. Nishizaki, and T. Fujita,
Phys. Rev. Lett. \textbf{76}, 3786 (1996).

\bibitem{Damascelli2000}
A. Damascelli et al.,
A. Damascelli, D. H. Lu, K. M. Shen, N. P. Armitage, 
F. Ronning, D. L. Feng, C. Kim, and Z.-X. Shen,T. Kimura and Y. Tokura,
Z. Q. Mao, and Y. Maeno
 Phys. Rev. Lett. \textbf{85}, 5194 (2000).

%\bibitem{Puchkov1998}
%A. V. Puchkov et al., Phys. Rev. B \textbf{58}, R13322, (1998).

\bibitem{Oguchi95}
T. Oguchi, Phys. Rev. B \textbf{51}, 1385 (1995).

\bibitem{Singh95}
D. J. Singh, Phys. Rev. B \textbf{52}, 1358 (1995).

\bibitem{Mazin99}
I. I. Mazin and D. J. Singh, Phys. Rev. Lett. \textbf{82}, 4324 (1999).

\bibitem{Sidis99}
Y. Sidis, M. Braden, P. Bourges, B. Hennion, S. NishiZaki, Y. Maeno, and Y. Mori,
Phys. Rev. Lett. \textbf{83}, 3320 (1999).


\bibitem{Zhitomirsky2001}
M. E. Zhitomirsky and T. M. Rice, Phys. Rev. Lett. \textbf{87}, 057001
(2001). 



\bibitem{Nomura2002}
T. Nomura and K. Yamada, J. Phys. Soc. Jpn. \textbf{71}, 404 (2002).


%\bibitem{Takimoto2000}
%T. Takimoto, Phys. Rev. B \textbf{62}, R14641 (2000).

%\bibitem{Agterberg97}
%D. F. Agterberg, T. M. Rice, and M. Sigrist, Phys. Rev. Lett. {\bf
%78}, 3374 (1997).

\bibitem{Sato2000}
M. Sato and M. Kohmoto, J. Phys. Soc. Jpn. \textbf{69}, 3505 (2000).


\bibitem{Kuwabara2000} 
T. Kuwabara and M. Ogata, Phys. Rev. Lett. \textbf{85}, 4586 (2000).


\bibitem{RossatMignod91}
J. Rossat-Mignod, L.P. Regnault, C. Vettier, P. Bourges,
P. Burlet, J. Rossy, J.Y. Henry and G. Lapertot,
Physica C \textbf{185-189}, 86 (1991).

\bibitem{Mook93}
H.A. Mook, M. Yethiraj, G. Aeppli, T.E. Mason and T. Armstrong,
Phys. Rev. Lett. \textbf{70}, 3490 (1993)

\bibitem{Fong95}
H.F. Fong, B. Keimer, P.W. Anderson, D. Reznik, F. Dogan and I. A. Aksay,
Phys. Rev. Lett. \textbf{75}, 316 (1995).

\bibitem{Bourges96}
P. Bourges, L.P. Regnault, Y. Sidis and C. Vettier,
Phys. Rev. B. \textbf{53}, 876 (1996).

\bibitem{Dai2001}
P. Dai, H.A. Mook, R. D. Hunt, and F. Dogan,
Phys. Rev. B \textbf{63}, 054525 (2001),



%\bibitem{Suhl59}
%H. Suhl, B. T. Matthias and L. R. Walker, Phys. Rev. Lett. \textbf{3},
%552 (1959).


\bibitem{Lu92} 
J.P. Lu, Phys. Rev. Lett. \textbf{68}, 125 (1992).

\bibitem{Bulut93}
N. Bulut and D. J.  Scalapino, Phys. Rev. B. \textbf{47}, 3419, (1993).

\bibitem{Tanamoto93}
T. Tanamoto, H. Kohno, and H. Fukuyama, 
J. Phys. Soc. Jpn. \textbf{62}, 1455 (1993).


\bibitem{Tanamoto94}
T. Tanamoto, H. Kohno, and H. Fukuyama, 
J. Phys. Soc. Jpn. \textbf{63}, 2739 (1994).

\bibitem{Marsiglio93}
F. Marsiglio, Phys. Rev. B {\bf 47}, 11555 (1993).

\bibitem{Lavagna 94}
M. Lavagna and G. Stemmann, 
Phys. Rev. B \textbf{49}, 4235 (1994).

\bibitem{Maki94}
K. Maki and H. Won, 
Phys. Rev. Lett. \textbf{72}, 1758 (1994).

\bibitem{Demler95}
E. Demler and S. C. Zhang,
Phys. Rev. Lett.  \textbf{75}, 4126 (1995).

\bibitem{Liu95}
D.Z. Liu and Y. Zha and K. Levin,
Phys. Rev. Lett. \textbf{75}, 4130 (1995).


\bibitem{Mazin95}
I.I. Mazin and V.M. Yakovenko,
Phys. Rev. Lett. \textbf{75}, 4134 (1995).


\bibitem{Brinckmann99}
J. Brinckmann and P.A. Lee,
Phys. Rev. Lett. \textbf{82}, 2915 (1999).


\bibitem{Brinkman74}
W.F. Brinkman, J. W. Serene, and P. W. Anderson, Phys. Rev. A \textbf{10},
2386 (1974).



\bibitem{Joynt88}
R. Joynt and T.M. Rice, Phys. Rev. B \textbf{38}, 2345 (1988).



\bibitem{Morr2001}
D. K. Morr, P. F. Trautman, and M. J. Graf, 
Phys. Rev. Lett. \textbf{86},
5978 (2001).


\bibitem{Kee2000}
H. Y. Kee, J. Phys. Condens. Matter \textbf{12}, 2279 (2000).

\bibitem{Fay2000}
D. Fay and L. Tewordt, Phys. Rev. B \textbf{62}, 4036 (2000).

\bibitem{Servant2002}
F. Servant et al., Phys. Rev. B \textbf{65}, 184511 (2002).


\bibitem{Braden2002}
M. Braden et al. Phys. Rev. B. \textbf{66}, 064522 (2002).


\bibitem{Lavagna94} 
M. Lavagna and G. Stemmann, Phys. Rev. B \textbf{49}, 4235 (1994).

\bibitem{Marsiglio93a}
F. Marsiglio, Phys. Rev. B {\bf 47}, 5419 (1993).


\bibitem{Mukuda2002}
H. Mukuda, K. Ishida, Y. Kitaoka, K. Miyake, Z. Q. Mao, Y. Mori, and Y. Maeno
Phys. Rev. B 65, 132507 (2002) 

\end{thebibliography}
\end{document}